\documentstyle[11pt,newpasp,twoside,epsf]{article}
\def\@journalname{Highlights of Astronomy}
\def\cpr@holder{Astronomical Society of the Pacific}
\def\@jourvol{$12$}
\def\cpr@year{2000}
\def\vol@title{JD9: Cold Gas and Dust at High Redshift}
\def\vol@author{D.J. Wilner, ed.}
\newcommand{\etal}{et~al.}

\def\edcomment#1{\iffalse\marginpar{\raggedright\sl#1\/}\else\relax\fi}
\reversemarginpar

\begin{document}
\title{Hidden Star Formation: The Ultraviolet Perspective}
 \author{G.R.\ Meurer, T.M.\ Heckman, M.\ Seibert}
\affil{The Johns Hopkins University, Department of Physics and Astronomy}
\author{J.\ Goldader}
\affil{University of Pennsylvania}
\author{D.\ Calzetti}
\affil{Space Telescope Science Institute}
\author{D.\ Sanders}
\affil{Institute for Astronomy, University of Hawaii}
\author{C.C.\ Steidel}
\affil{California Institute of Technology}

\begin{abstract}
Many recent estimates of the star formation rate density at high
redshift rely on rest-frame ultraviolet (UV) data.  These are highly
sensitive to dust absorption.  Applying a correlation between the
far-infrared (FIR) to UV flux ratio and UV color found in a local
starbursts to galaxy samples out to $z \sim 3$, one can account for most
of the FIR background.  However, the correlation is based on a sample
that does not include the most extreme starbursts, Ultra Luminous
Infrared Galaxies (ULIGs).  Our new UV images of ULIGs show
that their FIR fluxes are underpredicted by this correlation by factors
ranging from 7 to 70.  We discuss how ULIGs compare to the various
types of high-$z$ galaxies: sub-mm sources, Lyman Break Galaxies, and
Extremely Red Objects.
\end{abstract}

\section{Why observe star forming galaxies in the ultraviolet?}

About 60\%\ of the {\em intrinsic} bolometric luminosity of 
star forming populations is emitted between 912\AA\ and 3000\AA\ with
little variation in this fraction with star formation duration (from
models of Leitherer \etal\ 1999).  Hence UV light is potentially very
useful for measuring star formation.  It represents direct emission from
hot main-sequence stars, the same stars that will provide the majority
of the mechanical energy feedback into the ISM.  The UV spectrum is very
rich in features that can be used to diagnose the stellar populations
and intervening ISM.  The overall intrinsic spectral slope in the UV is
fairly constant.  For young ionizing populations it is set by the
Rayleigh-Jeans tail of the Planck function.  When looking at high
redshifts the importance of the rest-frame UV increases; at a
fixed bandpass it is the last stop before the Lyman-edge.

The big problem with using UV measurements is dust, which most
efficiently absorbs and scatters UV radiation.  The absorbed UV light is
reradiated at far infrared (FIR) wavelengths.  Not only is the amount of
dust important, so to is its distribution.  If the geometry is
unfavorable, as in the mixed stars and dust model, then the
UV emission will be dominated by the stars closest to the observer and
the center may be practically invisible (e.g.\ Witt, Thronson, \&\ Capuano
1992).  Fortunately, this is not the case for a
wide range of starburst galaxies as shown in Fig.~1.  For them the FIR
to UV flux ratio, or infrared excess (IRX), which gives the effective
dust absorption, correlates with the UV spectral slope $\beta$ (Meurer
\etal\ 1995, 1999).  This is a prediction of the dust screen model (Witt
\etal).  While variations on this geometry can also
explain this correlation (Calzetti 1997; Charlot \&\ Fall 2000), the
mere fact of this correlation allows one to recover the intrinsic
luminosity of starbursts using UV quantities alone.

\section{Applying IRX-$\beta$ to the high-redshift universe}

Lyman Break Galaxies (LBGs) at $z \sim 3$ are similar to local
starbursts. In particular they have similar SEDs (Dickinson 2000),
spectral properties (Tremonti \etal\ 2000), ISM dynamics (e.g.\ Pettini
\etal\ 1999), and surface brightnesses (Meurer \etal\ 1997).  They are
also noticeably redder than dust-free starbursts.  Meurer \etal\
(1999) used the rest frame colors of LBGs and the IRX-$\beta$
correlation to estimate that the Hubble Deep Field LBG sample suffers
from a factor of about five in dust absorption at rest $\lambda_0 \approx
1600$\AA.  It is hard to test whether the IRX-$\beta$ correlation
actually holds for LBGs because predicted FIR fluxes are typically just
below current detection limits with instruments such as SCUBA.  Other data
show that what little we know about the LBGs is consistent with them
obeying the same reddening law as local starbursts.  For instance,
we assumed that the FIR-radio correlation holds, and took 
the LBGs with the top ten predicted FIR emission and predicted
summed radio fluxes of 27 $\pm 5 \, \mu$Jy, and 105 $\pm 24 \, \mu$Jy
(assuming a 0.3 dex uncertainty on each flux) at
observed frame wavelengths 3.5cm and 20cm, while the Richards (2000) data yield
measured summed fluxes of 28 $\pm 10 \, \mu$Jy, and 100 $\pm 33 \,
\mu$Jy respectively.  Adopting IRX-$\beta$, Adelberger \&\ Steidel
(2000) show that the UV detectable galaxies at $z = 1, 2$ and 3 can
account for most or all of the FIR background at 850 $\mu$m.  

This rosy view of the utility of UV astronomy flies in the face of what
we have learned over the last few decades: that dust enshrouded star
formation is best seen in the infrared.  Would not dust obstruct
our view of star formation, even at high-$z$?  The bright SCUBA sources,
in particular, are inferred to have $z \approx 1 - 3$, usually have
little or no rest-frame UV emission and probably have an equal
contribution to the star formation rate density as non-dust corrected
LBGs (Barger, Cowie, \&\ Richards 2000).  They have $L_{\rm bol} >
10^{12}\, L_\odot$, so the best local analog to them are thought to be
the Ultra-Luminous Infrared Galaxies (ULIGs).  Relatively little was
known about the UV properties of ULIGs, until Trentham, Kormendy \&\ Sanders (1999)
presented weak UV detections of three ULIGs with HST's {\em Faint Object
Camera\/} and Surace \&\ Sanders (2000) showed that they are detectable
from the ground in the $U'$ band.

\begin{figure}
\plottwo{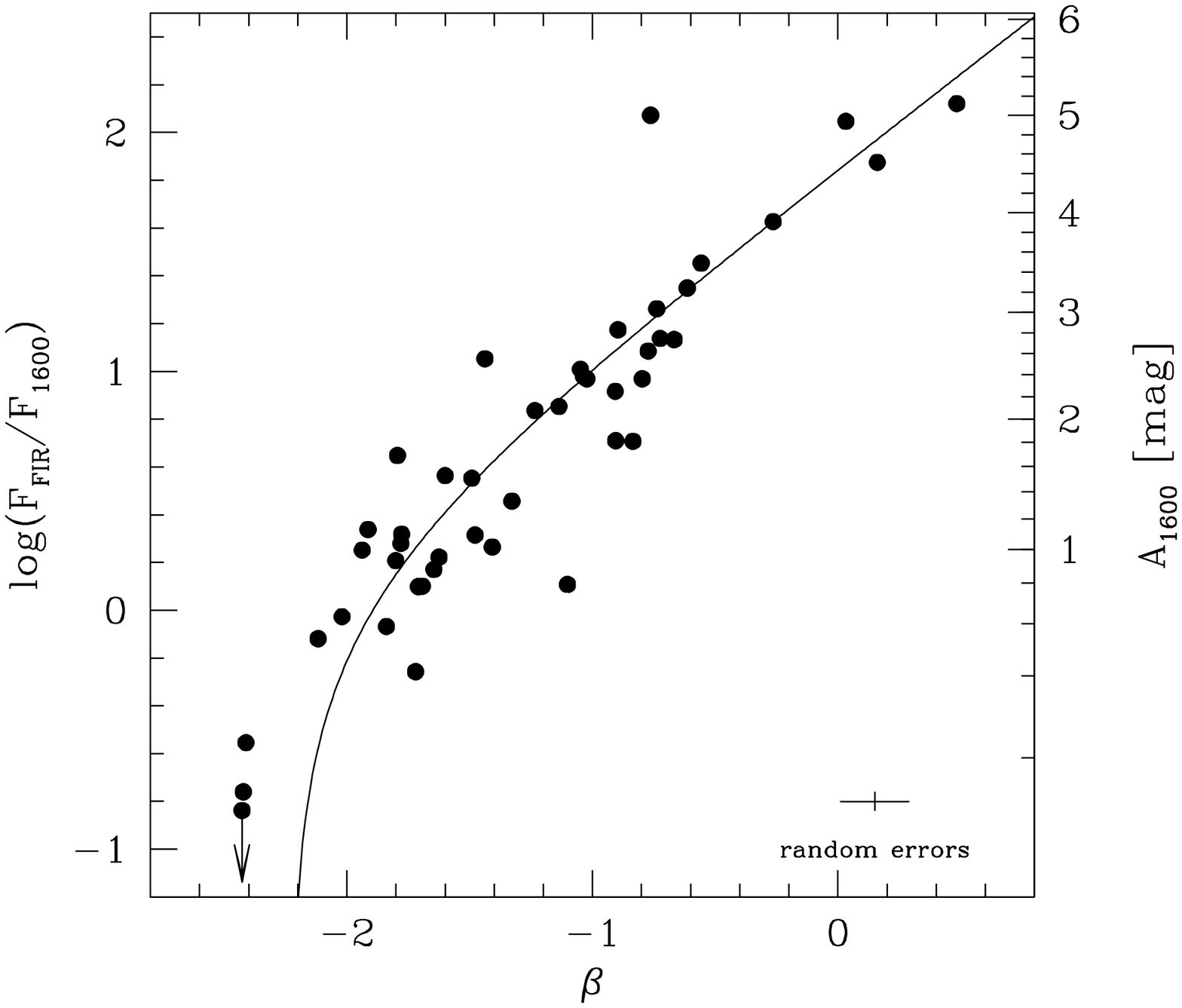}{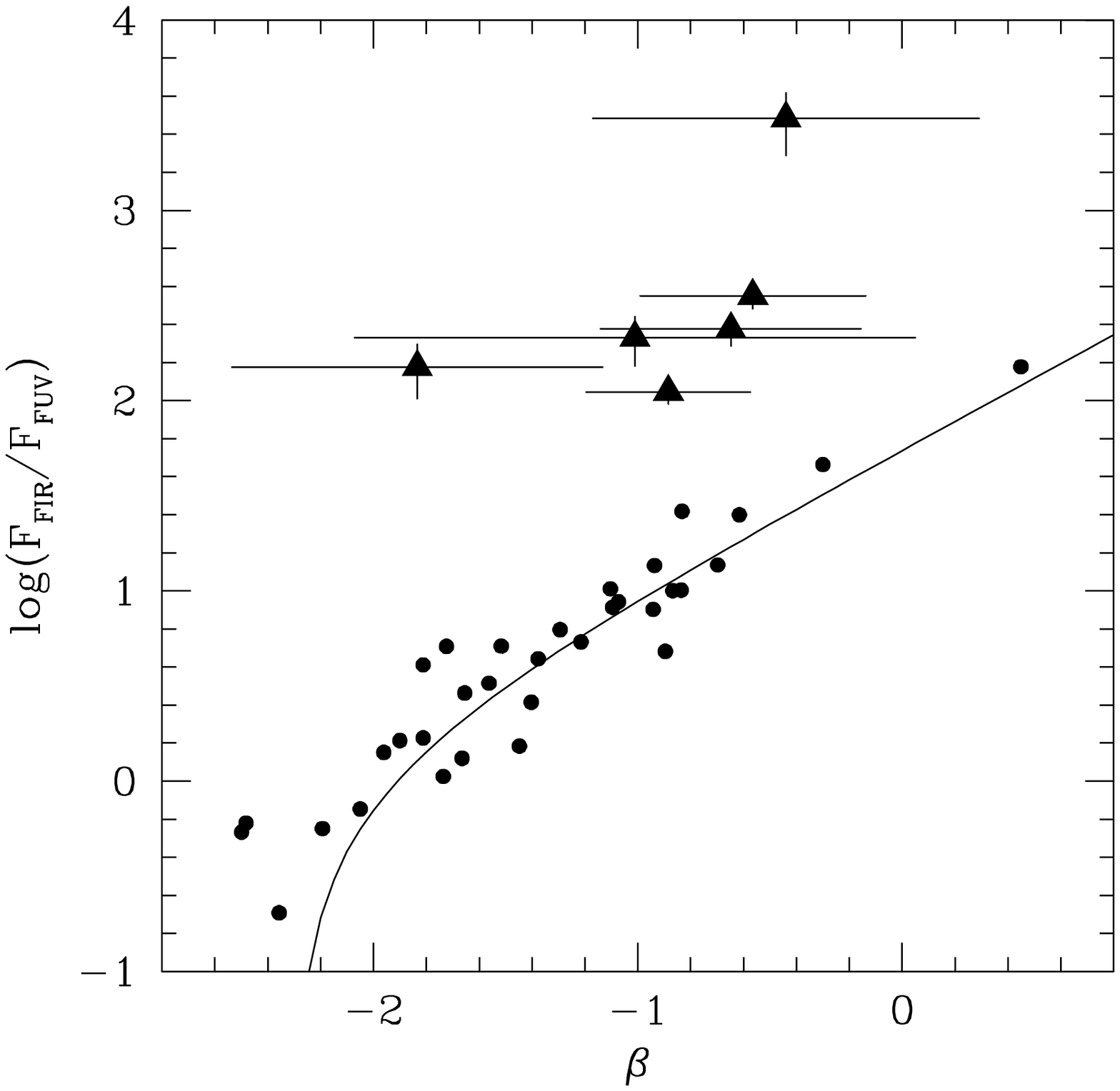}
\caption{(Left) The $IRX \equiv F_{\rm FIR}/F_{\rm UV}$
versus $\beta$ relationship of local starbursts observed by IUE (Meurer
\etal\ 1999).  Here the UV flux is measured at $\lambda_0 = 1600\AA$,
and the left axis converts IRX to the effective absorption $A_{1600}$ in
magnitudes. \\ Figure 2.\hskip 1.5em  (Right)  The IRX-$\beta$ plot for ULIGs
(triangles) in relation to the IUE starbursts.  Here $\beta$ and the UV flux
are measured photometrically through the actual or
synthetic STIS bandpasses. }
\end{figure}

\section{STIS UV Images of ULIGs}

We have been granted HST time to image seven galaxies in the UV using
the STIS ({\em Space Telescope Imaging Spectrograph\/}) MAMAs which have
much higher sensitivity than the FOC.  Our sample was chosen to have
$\log(L_{\rm bol}/L_\odot) \ge 11.6$, starting in $L_{\rm bol}$ where
the IUE sample ends.  So far six galaxies have been observed, five of
these are ULIGs.  We detect all of these in both the far UV
(FUV; $\lambda_c \approx 1460$\AA) and near UV (NUV; $\lambda_c \approx
2350$\AA).  In both bands, UV emission can be detected projected to
within 1 Kpc of the infrared nuclei seen by NICMOS (Scoville
\etal\ 2000).  However, especially in the FUV, very little, if any UV
emission is detected within the inner few hundred parsecs where most of
the bolometric luminosity probably originates.

Figure 2 shows the IRX-$\beta$ diagram for our sample compared to the
IUE starbursts.  Typically only 0.5\%\ of the bolometric luminosity is
observed in the FUV.  Furthermore, the IRX-$\beta$ correlation {\em
under-predicts\/} the FIR emission of ULIGs by factors ranging from 7 to
70.  The FIR flux is still under-predicted if only the light within 1
Kpc of the IR nucleus is considered.  In these galaxies IRX-$\beta$ only
gives a lower limit to the FIR flux.  These results confirm that ULIGs
represent galaxies with star formation almost totally hidden from the
UV.

\section{High-$z$ Implications}


Our work shows that we must still be cautious with rest-frame UV
observations of galaxies: they may harbor hidden-star formation beyond
that predicted with the IRX-$\beta$ correlation.  However, not all
galaxies are as extreme as ULIGs.  At high-$z$, the most luminous LBGs
can not generally have IRX values like ULIGs or else more would be
detected at 850$\mu$m with SCUBA (multiply predictions in Table 4 of
Meurer \etal\ 1999 by 7 -- 70).
Could ULIGs be selected as LBGs? At $z = 3$ ULIGs would have an observed
frame $V-I \leq 0.5$ ABmag.  Presumably they would have very red $U-B$
colors resulting from the strong opacity of the Lyman forest and edge.
Therefore, they should have the right colors to be selected as LBGs.
However at this redshift they would be very faint, having $V \sim 27$ to
30 ABmag, at or beyond the limits of many current surveys such as the
Hubble Deep Fields.  So ULIGs still make good analogs to SCUBA sources,
but probably only contribute to the faint end of the LBG population.

ULIGs have also been touted as good local prototypes for Extremely Red
Objects (EROs). Using our photometry and published results we find that
ULIGs emit enough rest-frame UV emission, that at $z=1.8$ or 3.5 in the
observed frame they would have $2.3 \leq R-K \leq 5.6$ ABmag.  This is
bluer than $R-K < 6$, the definition of EROs (Graham \&\ Dey 1996).
Since our sample is small, and some approach this limit perhaps some
more extreme ULIGs may be recognized as EROs at high-$z$, but in
general they would be too blue.

\end{document}